\documentclass[letterpaper]{JHEP3}

\usepackage{epsfig}
\usepackage{graphicx}

\newcommand{\labell}[1]{\label{#1}} 
\newcommand{\reef}[1]{(\ref{#1})}
\newcommand{\ie}{{\it i.e.,}\ }
\newcommand{\eg}{{\it e.g.,}\ }

\def\IR{{\hbox{{\rm I}\kern-.2em\hbox{\rm R}}}}
\newcommand{\be}{\begin{equation}}
\newcommand{\ee}{\end{equation}}
\newcommand{\beq}{\begin{equation}}
\newcommand{\eeq}{\end{equation}}
\newcommand{\beqa}{\begin{eqnarray}}
\newcommand{\eeqa}{\end{eqnarray}}
\newcommand{\prt}{\partial}

\newcommand{\lc}{\ell}
\newcommand{\tlc}{\tilde{\lc}}
\newcommand{\cF}{{\cal F}}
\newcommand{\tphi}{\tilde{\phi}}

\preprint{{\small hep-th/0211025}} \keywords{dS/CFT correspondence, field theory dualities}

\title{Tall tales from de Sitter space II:\\Field theory dualities}
\author{Fr\' ed\' eric Leblond\footnote{E-mail: {\tt
fleblond@hep.physics.mcgill.ca}},
Donald Marolf\footnote{E-mail: {\tt marolf@physics.syr.edu}} and
Robert C. Myers\footnote{E-mail: {\tt rmyers@perimeterinstitute.ca}}\\
$^{*,\ddagger}$ Perimeter Institute for Theoretical Physics,
Waterloo, Ontario N2J 2W9 Canada\\
$^{\ \ \dagger}$ Physics Department, Syracuse University, Syracuse, New York
13244 USA\\
$^{*,\ddagger}$ Department of Physics, McGill University, Montr\'eal,
Qu\' ebec H3A 2T8 Canada\\
$^{\ \ \ddagger}$ Department of Physics, University of Waterloo,
Waterloo, Ontario N2L 3G1 Canada}
\date{December, 2001}
\abstract{We consider the evolution of massive scalar fields in
(asymptotically) de Sitter spacetimes of arbitrary dimension.
Through the proposed dS/CFT correspondence, our analysis points to
the existence of new nonlocal dualities for the Euclidean
conformal field theory. A massless conformally coupled scalar
field provides an example where the analysis is easily explicitly
extended to `tall' background spacetimes.}

\begin{document}
\setcounter{footnote}{0}

\section{Introduction}
\label{intro}

Observations\cite{supernova} suggest that our universe
is proceeding towards a phase where its evolution is dominated by
a small positive cosmological constant. This suggestion poses new
challenges for string theory, which has seen much success in
asymptotically flat spaces and in settings with an effective
negative cosmological constant (\eg Freund-Rubin
compactifications\cite{RubF} producing asymptotically anti-de
Sitter spaces). The impressive success of the AdS/CFT
correspondence \cite{adsrev}, which has provided fairly concrete
realizations of a holographic duality between quantum gravity and
a field theory in one dimension lower without gravity, has
prompted speculations that it may be possible to describe string
theory or quantum gravity in asymptotically de Sitter spaces by a
similar dS/CFT correspondence \cite{AS1,witten}.  Like the AdS
case, the symmetries of de Sitter space suggest that the dual
field theory is conformally invariant.

Of course, the nature of de Sitter space is quite different from
its AdS counterpart.  In particular the conformal boundaries,
which one expects to play a central role in any dS/CFT
correspondence, are hypersurfaces of Euclidean signature.  As a
result, one expects the dual field theory to be a Euclidean field
theory.  Further in de Sitter space, there are two such
hypersurfaces: the future boundary, $I^+$, and the past boundary,
$I^-$.  Hence one must ask whether the proposed duality will
involve a single field theory \cite{AS1,witten} or two
\cite{annals}. Unfortunately at present, the most striking
difference from the AdS/CFT duality is the fact that we have no
rigorous realizations of the dS/CFT duality.

However, the idea of a dS/CFT correspondence is a powerful and
suggestive one that could have fundamental implications for the
physics of our universe.  A present difficulty is that rather
little is known about the Euclidean field theory which is to be
dual to physics in the bulk.  Our goal here is to explore further
the requirements for a candidate dual field theory under the
assumption that the bulk physics must reproduce standard
background quantum field theory in the low energy limit.  In
particular then, up to possible quantum gravity violations, it
should be possible to express the observables\footnote{Here we use
the term `observables' in the technical sense of `gauge invariant
operators' without direct concern for a sense in which these
observables might be measurable \cite{Preskill,Sorkin}.} localized
near any Cauchy surface in terms of the  observables localized
near any other Cauchy surface. Of course, this is what one usually
refers to as `unitarity,' though even in standard quantum field
theory it may not represent unitary evolution in the technical
sense \cite{TV,WaldQFT}. Below, we will find that this benign
assumption about the bulk physics has rather extraordinary
consequences for the dual CFT.

Our primary tool for exploring these consequences is Einstein gravity
with a positive cosmological constant coupled to an otherwise free scalar
field. For simplicity, we consider quantum field theory for the scalar
linearized about fixed de Sitter or asymptotically de Sitter
backgrounds.  However, it will be clear that, at least as far
as the above property is concerned, linearized gravitons and
even non-linear background quantum field theory must yield similar
results. Before proceeding with a detailed discussion of the scalar
field, let us briefly present our main observations.

Central to our investigations are the two conformal boundaries
appearing in de Sitter space, and the claim \cite{AS1,witten} that
only a single dual field theory is needed. Recently, an alternate
point of view has been advocated \cite{annals} in which the dual
description of dS space involves an entangled state of two field
theories associated with the two separate boundaries. We will
comment on this idea in the discussion (section \ref{disc}), but
our present focus will be on the single CFT proposal where one
does not have a `separate' dual theory associated with $I^+$ and
with $I^-$. Some intuition for this point of view can be found by
combining the above assumption of bulk unitarity with the central
postulate of the dS/CFT correspondence. That is, operators in the
dual field theory may be associated with appropriate limits of
bulk operators pushed to either $I^-$ or $I^+$. So consider the
full set of bulk operators localized near some arbitrary Cauchy
surface. Now unitary time evolution in the bulk allows us to
`push' this complete set of operators to either $I^-$ or $I^+$ and
hence obtain corresponding sets of dual operators. Of course,
since we began with a complete set of bulk operators, correlation
functions for any bulk operators can then be computed in terms of
those of the dual operators on either boundary. In particular,
correlation functions in the dual theory associated with $I^+$ can
be expressed in terms of correlation functions associated with
$I^-$ and vice versa.

The original discussion of \cite{AS1,witten} emphasized the
causal connection between points on the two boundaries of dS space.
In particular, a light cone emerging from a point on $I^-$ expands into
the space and reconverges at the antipodal point on the sphere at $I^+$.
As a consequence, the singularity structure of certain boundary correlation
functions is left invariant when, \eg a local operator on $I^-$ is
replaced by a corresponding local operator at the antipodal point
on $I^+$ \cite{AS1}, as will be reviewed in section \ref{nonlocal}. Not
only does this observation suggest that there is a single dual
field theory, but further that dual operators associated with the
two boundaries are simply related by the antipodal map on the
sphere. The latter would be surprising as one expects that in
general the time evolution map connecting the corresponding bulk
operators is nonlocal. In fact, as will be discussed below, this
latter intuition is correct. The mapping between dual operators on
$I^\pm$ is highly nonlocal, as is readily revealed by explicitly
examining the (retarded) Green's functions in dS space. While the
singularities in the Green's function propagate along the light
cone, generically there is also nontrivial support within the
light cone. For certain special cases, however, the time evolution
map does produce a simply antipodal mapping between $I^+$ and
$I^-$. Given the lack of guidance coming from a working example of
the proposed duality, one might interpret this result as a hint
towards the specific types of fields that would appear in a
successful realization of the dS/CFT. Unfortunately, however, this
selection rule based on locality of the mapping between boundaries
does not seem to bear up in more interesting applications, as
follows. \FIGURE{\epsfig{file=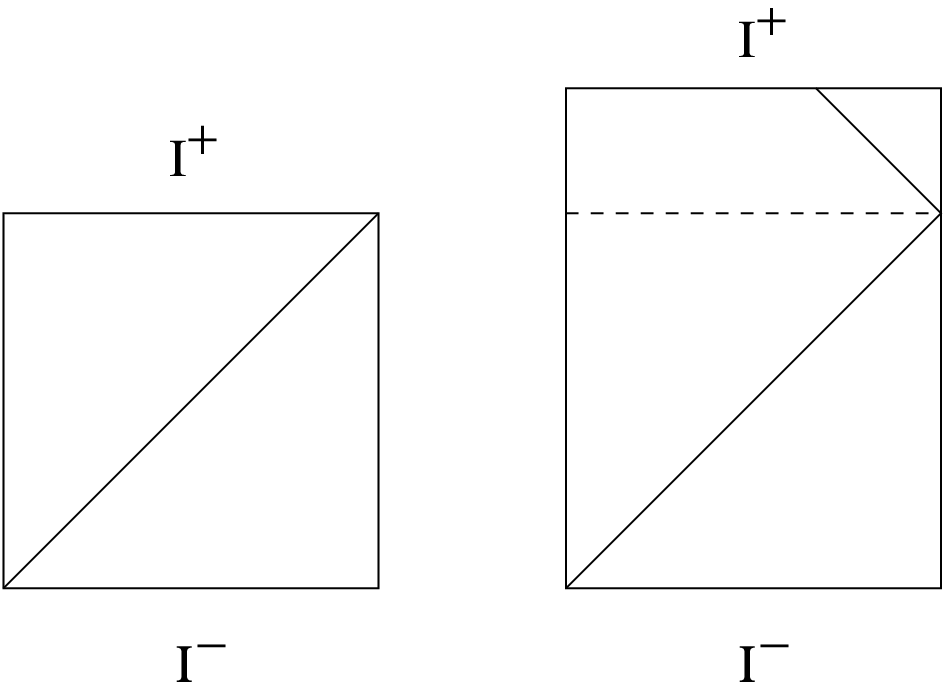, width =
7cm}\caption{Conformal diagrams of a) de Sitter space and  b)
perturbed de Sitter space.} \label{heights}}

The discussion of the dS/CFT correspondence has also been extended
beyond pure dS space to more general backgrounds with
asymptotically dS regions. Again, in analogy to the AdS/CFT
duality, such backgrounds might have an interpretation in terms of
`renormalization group flows' in the dual field theory
\cite{AS,BBM,tale1,rgflows,rgflows2}. Now it is a straightforward
consequence of a theorem of Gao and Wald \cite{GW} that in such a
generic (nonsingular) background, observers are able to view an
entire Cauchy surface at a finite time. The corresponding
conformal diagrams may be described as `tall' (see figure
\ref{heights}--- see also \cite{tale1} for details).

The nonlocality in the relation between dual operators on the two
boundaries becomes manifest when we consider such `tall'
backgrounds. The above discussion, in which the bulk evolution map
relates the CFT operators associated with $I^+$ and $I^-$, remains
essentially unchanged. However, a key difference is that the
causal connection between $I^-$ and $I^+$ is now manifestly {\it
not} local. In a tall spacetime, the light rays emerging from a
point on $I^-$ reconverge,\footnote{For simplicity, our
description is restricted to spherically symmetric foliations
\cite{tale1}. Generically the converging light rays would not be
focussed to a single point.} but this occurs at a finite time long
before they reach $I^+$. After passing through the focal point,
the rays diverge again to enclose a finite region on $I^+$. This
observation precludes any intuition that the dual operators
associated with the two boundaries could be related by a simple
local map (\eg the antipodal map) on the sphere. That is,
following the arguments presented above, one is still lead to
conclude that in these tall backgrounds pushing the bulk operators
to $I^-$ and $I^+$ must yield equivalent dual physics, but the map
that implements this equivalence can no longer be local. Hence in
the context of the dS/CFT duality, it seems that nonlocality will
be an unavoidable aspect of the relation between field theory
operators associated with two conformal boundaries. One can
reproduce precisely the same boundary correlators or observables
but through some nonlocal reorganization of the degrees of freedom
within the dual field theory. It seems appropriate to refer to
such relations as {\it nonlocal dualities} within the field
theory.

The remainder of the paper is organized as follows: We begin with
a brief review of the dS/CFT correspondence in section \ref{gen}.
This also allows us to emphasize certain subtleties that arise in
this proposal, and it sets the stage for our main discussion. In
section \ref{nonlocal}, we explore in more detail the consequences
of the bulk evolution map for the relationship between the dual
field theories associated with the two boundaries, $I^-$ and
$I^+$. Section \ref{disc} provides a short discussion of our
results. Various useful technical details about scalar field
theory in de Sitter space are presented in appendices. Appendix
\ref{appendix1} provides a detailed analysis of the evolution of
massive scalar fields in a pure ($n$+1)-dimensional dS background,
while appendix \ref{appendix3} considers the stability of dS space
with respect to perturbations by the scalar field modes.

\section{Generalities of the dS/CFT correspondence}
\label{gen}

We wish to discuss the interpretation of scalar field theory in a
de Sitter background in the context of the dS/CFT correspondence.
However, much of the motivation for the dS/CFT duality, as well as
the interpretation of the dS space calculations, comes from our
understanding of the AdS/CFT duality \cite{adsrev}. Hence we begin
with a brief review of the latter correspondence (section
\ref{AdSrev}) and then present the most relevant aspects of dS/CFT
(section \ref{dSrev}). Much of this discussion is review, although
our emphasis is somewhat different than in previous treatments.
The reader interested in the mathematical details can consult the
appendices to which we will refer in the following.

\subsection{Brief AdS/CFT review}
\label{AdSrev}

Consider probing anti-de Sitter space with a massive scalar field.
We consider the following metric on ($n$+1)-dimensional AdS
space,\footnote{The essential feature for the following analysis
is the exponential expansion of the radial slices with proper
distance $r$. While we have chosen to consider pure AdS space in
Poincar\'e coordinates for specificity, this expansion, of course,
arises quite generally in the asymptotic large-radius region for
any choice of boundary metric and for any asymptotically AdS
spacetime.}
\beq \labell{metric1}
ds^2=dr^2+e^{2r/\tlc}\eta_{\mu\nu}dx^\mu dx^\nu\ ,
\eeq
and the standard equation of motion for the scalar,
\beq \labell{eomscale}
\left[ \Box - M^{2} \right] \phi = 0\ .
\eeq
Then to leading order
in the asymptotic region $r\rightarrow\infty$, the two independent
solutions take the form \cite{wittenads}
\beq \labell{asympscal}
\phi_\pm\simeq e^{-\Delta_\pm r/\tlc} \phi_{0\pm}(x^\mu)
\qquad{\rm where}\qquad
\Delta_\pm={n\over2}\pm\sqrt{{n^2\over4}+M^2\tlc^2}\ .
\eeq
Now the interpretation of these results depends on the value of the
mass, and there are three regimes of interest:
\beq\labell{threem}
{\rm (i)}\ \  M^2>0\ ,\ \ \ {\rm (ii)}\ \ 0>M^2>-{n^2\over4\tlc^2}\ \ \
{\rm and}\ \ \ {\rm (iii)}\ \ M^2<-{n^2\over4\tlc^2}\ .
\eeq

In case (i), $\Delta_-$ is negative and so the corresponding
``perturbation'' is actually divergent in the asymptotic regime.
Hence in constructing a quantum field theory on AdS space, only
the $\phi_+$ modes would be useful for the construction of an
orthogonal basis of normalizable mode functions \cite{vijay1}.  In
particular, the bulk scalar wave operator is essentially
self-adjoint and picks out the boundary condition that the
$\phi_-$ modes are not excited dynamically. In the context of the
AdS/CFT then, the $\phi_{0-}$ functions are associated with source
currents (of dimension $\Delta_-$). These may then be used to
generate correlation functions of the dual CFT operator of
dimension $\Delta_+$ through the equivalence
\cite{wittenads,steve}
\beq\labell{equiv}
Z_{\rm AdS}(\phi)=\int D\phi\, e^{iI_{\rm AdS}(\phi_-,\phi_+)}=
\left\langle e^{i\int \phi_{0-}\cal{O}_-}
\right\rangle_{\rm CFT} .
\eeq
 On the other hand, the boundary
functions $\phi_{0+}$ are associated with the expectation value
for states where the dual operator has been excited \cite{vijay1}.

In case (ii), the lower limit corresponds precisely to the
Breitenlohner-Freedman bound \cite{BF,paul}. While the scalar
appears tachyonic, it is not truly unstable and it is still
possible to construct a unitary quantum field theory on AdS space.
Further, in this regime, both sets of solutions \reef{asympscal}
are well-behaved in the asymptotic region.  However, together they
would form an over-complete set of modes. The theory must
therefore be supplemented with a boundary condition at AdS
infinity which selects out one set of modes to define a
self-adjoint extension of the scalar wave operator (and thus the
time evolution operator). For $0>M^2\tlc^2>1-n^2/4$, there is a
unique boundary condition which produces an AdS invariant
quantization \cite{BF}. However, for \beq \labell{ambiguous}
1-n^2/4>M^2\tlc^2>-n^2/4\ , \eeq the boundary condition is
ambiguous. The AdS/CFT interpretation is essentially the same as
above. That is, the $\phi_{0+}$ and $\phi_{0-}$ functions may be
associated with expectation values and source currents of the dual
CFT operator, respectively. For the ambiguous regime
\reef{ambiguous}, there is a freedom in this equivalence
associated with a Legendre transformation of the generating
functional \cite{igor1}.

Finally in case (iii), the mass exceeds the Breitenlohner-Freedman bound
\cite{BF,paul} and the scalar field is actually unstable;
no sensible quantization is possible. However, if one were to attempt an
AdS/CFT interpretation analogous to those above, the dimension $\Delta_+$ of the
dual CFT operator would be complex, which might be interpreted as indicating
that the corresponding theory is not unitary. Hence one still seems to have
agreement on both sides of the correspondence as to the unsuitability
of the regime $M^2 \tilde \ell^2 < -n^2/4$.

\subsection{Some dS/CFT basics}
\label{dSrev}

Given the brief overview of the AdS/CFT correspondence, we now
turn to asymptotically de Sitter spaces, where one would like to
study the possibility of a similar duality between quantum gravity
in the bulk and a Euclidean CFT \cite{AS1}. As in the previous
review, we focus the present discussion on the case of a pure de
Sitter space background:
\beq \labell{metricds}
ds^{2} = -dt^{2}
+\cosh^{2}t/\lc\ d\Omega^{2}_{n}\ ,
\eeq
where $d\Omega^2_n$ is the standard round metric on an $n$-sphere.
This metric solves Einstein's equations,
$R_{ij}=2\Lambda/(n-1)\,g_{ij}$, in $n$+1 dimensions. The
curvature scale $\lc$ is related to the cosmological constant by
$\lc^2=n(n-1)/(2\Lambda)$. Again the important feature of this
geometry is the exponential expansion in the spatial metric in the
asymptotic regions, \ie $t\rightarrow\pm\infty$. Much of the
following discussion carries over to spacetimes that only resemble
dS asymptotically\footnote{Many explicit examples of such
backgrounds may be found in ref.~\cite{tale1}.} and indeed, if the
proposed dS/CFT duality is to be useful, it must extend to such
spacetimes. We will explore certain aspects of the dS/CFT for such
backgrounds in section \ref{nonlocal}.

Consider a free scalar field propagating on the above background
\reef{metricds}, which we wish to treat in a perturbative regime
where the self-gravity is small. Hence, the equation of motion is
\beq \labell{eomfirst2} \left[ \Box - M^{2} \right] \phi = 0\ .
\eeq In general, the effective mass may receive a contribution
from a nonminimal coupling to the gravitational field \cite{BD}.
Therefore we write
\beq
\labell{messs}
M^{2} = m^{2} + \xi R\ ,
\eeq
where $m^{2}$ is the mass squared of the field in the flat space
limit and $\xi$ is the dimensionless constant determining the
scalar field's coupling to the Ricci scalar, $R$. In the dS
background \reef{metricds}, we have $R=n(n+1)/\lc^2$. A case of
particular interest in the following section will be that of the
conformally coupled massless scalar field, for which $m^2=0$,
$\xi=(n-1)/4n$ and hence $M^2=(n^2-1)/4\lc^2$. With these
parameters, the solutions of eq.~\reef{eomfirst2} transform in a
simple way under local conformal scalings of the background metric
\cite{BD}.

In parallel with the AdS case, scalar fields propagating in de
Sitter space can have two possible behaviors near the boundaries.
Let us for the moment think of defining these boundary conditions
at past infinity ($I^-$). Equation (\ref{eomfirst2}) above is
readily solved \cite{AS1} near $I^-$ to yield two independent
solutions with the asymptotic form  $\phi \sim e^{h_\pm t/\lc}$
where
\begin{equation}
\labell{exponents}
h_{\pm} = \frac{n}{2} \pm \sqrt{
\frac{n^{2}}{4} - M^{2}\ell^2} \ .
\end{equation}
Note that this asymptotic time dependence is independent of the
details of the spatial mode. In the pure dS background
\reef{metricds}, the same exponents also govern the behavior of
the fields at future infinity --- see appendix \ref{appendix1} for
details.

The fact that the boundaries are spacelike in de Sitter space
means that the `boundary conditions' have a different conceptual
status than in the AdS setting.  In particular, requiring that the
bulk evolution is well-defined in dS space will not impose any
restrictions on past or future boundary conditions. So in contrast
to the AdS/CFT correspondence, in the dS/CFT correspondence, both
the $\phi_+$ and $\phi_-$ modes appear on an equal
footing.\footnote{Attempts have been made to distinguish these
modes through energy considerations \cite{rgflows2} but we
disagree with their discussion, as described in appendix
\ref{appendix3}.} Certainly, a complete description of physics in
the bulk must include both sets of modes as dynamical quantum
fields. Following the analogy with the AdS/CFT correspondence and
in accord with the preceding discussion, it is natural then to
associate both modes $\phi_\pm$ with source currents for dual
field theory operators ${\cal O}_\pm$, with conformal dimensions
$h_{\mp}$ \cite{AS1}. As we will discuss shortly, this matching of
modes with dual operators is further supported by a bulk
construction of a generating functional for correlation functions
in the CFT.

As in the AdS case, one can classify the scalars displaying
distinct types of boundary behavior in three different regimes:
\beq\labell{threem2} {\rm (i)}\ \  M^2>{n^2\over4\lc^2}\ ,\ \ \
{\rm (ii)}\ \ {n^2\over4\lc^2}>M^2>0\ \ \ {\rm and}\ \ \ {\rm
(iii)}\ \ M^2<0\ . \eeq
These three regimes also appear in discussions in the mathematics
literature --- see, \eg \cite{tagirov,RLN,RLN2,RLN3}. There the
scalar field is classified according to $M^{2}$ regarded as its
SO($n+1,1$) Casimir. A common nomenclature for the three
possibilities delineated above is the (i) principal, (ii)
complementary (or supplementary) and (iii) discrete series of
representations of SO($n+1,1$). As is evident from
eq.~\reef{exponents}, the distinguishing feature of scalar fields
in the principal series is that they are oscillatory near past (or
future) infinity.  In contrast, the exponents for fields in the
complementary series are real and positive, and so their
asymptotic behavior is a purely exponential damping near both
boundaries.

Let us consider case (iii) $M^2<0$ in detail. While $h_{\pm}$ are
both real, the modes $\phi_- \sim e^{h_- t}$ diverge as one
approaches $I^-$ since $h_-<0$. One finds similar divergent
behavior for one of the modes at the future boundary $I^+$. The
discrete series then corresponds to special values of the mass in
this range where a subset of the modes display the convergent
$h_+$ behavior at both $I^\pm$ --- see Appendix \ref{appendix1}
and refs.~\cite{tagirov,RLN,RLN2,RLN3}. However, we emphasize that
even in these special cases, the full space of solutions still
includes modes diverging at both asymptotic boundaries. In a
physical situation then, the uncertainty principle would not allow
us to simply set the amplitude of the divergent modes to zero.
Hence the formal mathematical analysis of these fields is only of
limited physical interest and we will not consider them further in
the following. Of course, the divergence of the generic field
configuration is simply an indication that treating the tachyonic
fields as linearized perturbations is inappropriate. Nonlinear
field theories with potentials including unstable critical points
may play an important role in the dS/CFT correspondence, \eg in
constructing models of inflationary cosmology. The essential point
though is that one must study the full nonlinear evolution of such
fields including their backreaction on the spacetime geometry.

Considering the principal series, (i) $M^2>n^2/4\lc^2$, in more
detail, one expects to find a pair of dual operators ${\cal
O}_\pm$ with {\it complex} conformal dimensions $h_{\mp}$. Having
operators with a complex conformal weight suggests that the dual
CFT is nonunitary \cite{AS1}. We add the brief observation that,
since in the quantum field theory, the two sets of independent
modes $\phi_\pm$ correspond roughly to creation and annihilation
operators (positive and negative frequency modes) in the bulk
(see, \eg \cite{AS3,allen2,em1,allen}), the corresponding
operators in the dual theory should have nontrivial commutation
relations.

Finally we consider the complementary series, (ii)
$n^{2}/{4\lc^2}> M^2 \geq 0$. In this case, the time dependence
for both sets of modes is purely a real exponential decay near
both $I^\pm$. In the bulk, the two linearly independent solutions
may be chosen to be real, as is readily verified by explicit
computations --- see appendix \ref{appendix1}. Because the
$\phi_\pm$ solutions are real, they each have zero norm in the
usual Klein-Gordon inner product, while a nonvanishing inner
product arises from $(\phi_+,\phi_-)$. It follows that upon
quantization the corresponding operator coefficients are analogous
to position and momentum operators, rather than creation and
annihilation operators. That is, these degrees of freedom are
canonically conjugate. In any event, both types of modes are again
required to describe standard quantum field theory in the bulk.

As before, the dual CFT should contain a pair of operators ${\cal
O}_{\pm}$ dual to the $h_\mp$ modes.  In this case, the operators
have real conformal weights and must be distinct as their weights
are different. One can readily see that both ${\cal O}_{\pm}$ will
have local correlation functions.  One simply notes that the
corresponding source currents are obtained from the bulk scalar
field through
\beqa
\labell{def1}
{J}_- (\Omega) &\equiv& \lim_{t\rightarrow -\infty} e^{-h_-t/\lc} \phi(\Omega,t)
\ ,
\\
{J}_+ (\Omega) &\equiv& \lim_{t\rightarrow -\infty} e^{-h_+t/\lc} [\phi(\Omega,t)
- e^{h_-t/\lc} {J}_-(\Omega)]\ ,
\nonumber
\eeqa
where $\Omega$ denotes a point on the $n$-sphere. As these
constructions are local in position, their two-point functions
will also be local. Note that the above discussion of inner
products indicates that the operators ${\cal O}_\pm$ should have
nontrivial commutation relations with each other but vanishing
commutators amongst themselves.

Next we consider the generator of correlation functions in the
dual field theory. A natural construction proposed in \cite{AS1}
for a free bulk field theory is
\beq \labell{evolmap}
\cF=\lim_{t,t'\rightarrow-\infty} \int
d\Sigma^{\mu} d\Sigma'^{\nu}
\phi(x)\stackrel{\leftrightarrow}{\partial}_{\mu} G(x,x')
\stackrel{\leftrightarrow}{\partial}_{\nu}\phi(x')\ .
\eeq
In the original proposal of \cite{AS1}, $G(x,x')$ was chosen as
the Hadamard two-point function
\beq \labell{hadamard}
G(x,x')=\langle 0|\lbrace\phi(x),\phi(x')\rbrace|0\rangle
\eeq
in the Euclidean vacuum, which is symmetric in its arguments.
Generalizing this construction to other two-point functions was
considered in \cite{AS3,AS4}. These alternatives all provide
essentially the same short distance singularities discussed below.

One proceeds by evaluating the generating functional $\cF$. First
the boundary conditions \reef{exponents} at $I^-$ yield
\beq \labell{boundary-} \lim_{t \rightarrow -\infty}\phi(\Omega,t)
\simeq \phi_{0+}(\Omega)\, e^{h_{+}t}+\phi_{0-}(\Omega)\,
e^{h_{-}t} \ , \eeq
where we imagine that $M^2>0$ so that the above shows no divergent
behavior. Now the dS-invariant two-point function may also be
expanded in the limit that $t,t'\rightarrow-\infty$ with the
result being
\beq \labell{mapans1}
G(x,x') \simeq
c_{+}{e^{-h_+(t+t')}\over(w^{i}w^{\prime i}-1)^{h_+}}
+c_{-}{e^{-h_-(t+t')}\over(w^{i}w^{\prime i}-1)^{h_-}}
\ ,
\eeq
where $c_{+}$ and $c_{-}$ are constants and $w^i$ denote direction
cosines on $S^n$. Using the notation of \cite{AS4}, one has
$w^{1}=\cos\theta_{1}, \; w^{2}=\sin\theta_{1}\cos \theta_{2}, \;
..., \; w^{d}=\sin\theta_{1}\, \ldots \,
\sin\theta_{n-1}\sin\theta_{n}$. Note in particular that with this
choice of coordinates when the points on sphere coincide, one has
$w^iw^{\prime i}=1$, while for antipodal points, one has
$w^iw^{\prime i}=-1$. Taking into account the measure factors, the
final result for the generating functional reduces to
\beq \labell{mapans2} \cF=-\frac{(h_+-h_-)^{2}}{2^{2n}} \int
d\Omega d\Omega'
\left[c_+\frac{\phi_{0-}(\Omega)\,\phi_{0-}(\Omega')}{(w^{i}w^{\prime
i}-1)^{h_{+}}}
+c_-\frac{\phi_{0+}(\Omega)\,\phi_{0+}(\Omega')}{(w^{i}w^{\prime
i}-1)^{h_{-}}} \right]\ . \eeq
Note that the Klein-Gordon inner product has eliminated the
cross-terms (which were potentially divergent). Further the
coincidence singularities in eq.~\reef{mapans2} are proportional
to the Euclidean two-point function on a $n$-sphere, \ie
\beq \labell{euctwo} \Delta_{h_{\pm}} \simeq
\frac{1}{(w^{i}w^{\prime i}-1)^{h_{\pm}}} , \eeq
for operators with conformal weight $h_\pm$. Hence $\cF$ appears
to be a generating functional for CFT correlation functions with
$\phi_{0\pm}$ acting as source currents for operators with
conformal dimensions $h_\mp$. The above relies on having a free
field theory in the bulk dS space, but extending the construction
to an interacting field theory was considered in \cite{AS4}.

\section{CFT on two boundaries}
\label{nonlocal}

As remarked in the introduction, de Sitter space has two conformal
boundaries and so one may ask the question as to whether the
dS/CFT correspondence involves a single dual field theory or two.
One simple argument in favor of one CFT is as follows \cite{ASC}:
The isometry group of ($n$+1)-dimensional dS space is SO($n+1,1$),
which agrees with the symmetries of a single Euclidean CFT in $n$
dimensions. Further note that the global Killing vector fields
corresponding to these isometries in dS space act nontrivially on
both $I^\pm$. Hence there is a simple correlated action on source
currents or dual operators identified with each of the boundaries.
Hence given the single symmetry group, it is natural to think that
the dual description involves a single CFT.

Further we would recall our experience from the AdS/CFT
correspondence. A central point in this context is that the CFT
does not `live' on the boundary of the AdS space. Usually one has
chosen a particular foliation of AdS \cite{us}, and the bulk space
calculations are naturally compared to those for the field theory
living on the geometry of the surfaces comprising this foliation.
Via the UV/IR correspondence, each surface in the bulk foliation
is naturally associated with degrees of freedom in the CFT at a
particular energy scale \cite{folia}. The boundary of AdS space
plays a special role in calculations as this is a region of the
geometry where the separation between operator insertions and
expectation values is particularly simple. One notable exception
where two CFT's seem to play a role is the eternal black hole
\cite{Juan,other,LM}. In this case, however, the bulk geometry has
two causally disconnected boundaries. In fact, one can show that
for any solution of Einstein's equations with more than one
asymptotically AdS boundary, the boundaries are all causally
disconnected from each other \cite{math}. In the case of dS space,
the past and future boundaries are certainly causally connected
and so it seems $I^\pm$ can be considered as two (special) slices
in a certain foliation \reef{metricds} of the spacetime. Hence
this reasoning suggests that one should only consider a single CFT
in the dual description.

\subsection{Nonlocality in the boundary map}
\label{whoopee}

Next we turn to Strominger's observation \cite{AS1} that the
generating functional \reef{evolmap} can in certain circumstances
be extended to incorporate sources on $I^+$. Certainly the
construction of the generating functional in the previous section
produces essentially the same result if we replace both of the
limits in eq.~\reef{evolmap} with $t,t'\rightarrow+\infty$. This
would produce an analogous generating functional with source
currents defined by the asymptotic behavior of the scalar near
$I^+$, \ie
\beq \labell{boundary+} \
\lim_{t \rightarrow +\infty}\phi(\Omega,t)
\simeq \tphi_{0+}(\Omega)\, e^{-h_{+}t}+\tphi_{0-}(\Omega)
\,e^{-h_{-}t} \ ,
\eeq
However, it is also interesting to consider the case where only
one of the limits in eq.~\reef{evolmap} is replaced with one
approaching $I^+$,
\beq \labell{evolmap2}
\widetilde\cF=\lim_{t\rightarrow+\infty,t'\rightarrow-\infty} \int
d\Sigma^{\mu} d\Sigma'^{\nu}
\phi(x)\stackrel{\leftrightarrow}{\partial}_{\mu} G(x,x')
\stackrel{\leftrightarrow}{\partial}_{\nu}\phi(x')\ .
\eeq
Now an essential observation \cite{AS1,witten} is the causal
connection between points on the two boundaries $I^\pm$. In
particular, a null geodesics emerging from a point on $I^-$
expands out into the dS spacetime and refocuses precisely at the
antipodal point on the $n$-sphere at $I^+$. Hence the two-point
function in eq.~\reef{evolmap2} (or any dS-invariant Green's
function) will introduce singularities when the point on $I^+$
approaches the antipode to the point on $I^-$, as the proper
separation of these points vanishes. In fact, in certain
circumstances (see the details below), evaluating the above
expression yields the simple result:
\beq \labell{mapans99} \widetilde\cF=-\frac{(h_+-h_-)^{2}}{2^{2n}}
\int d\Omega d\Omega' \left[\tilde{c}_+
\frac{\tphi_{0-}(\Omega)\,\phi_{0-}(\Omega')}{(w^{i}w^{\prime
i}+1)^{h_{+}}}
+\tilde{c}_-\frac{\tphi_{0+}(\Omega)\,\phi_{0+}(\Omega')}{(w^{i}w^{\prime
i}+1)^{h_{-}}} \right]\ . \eeq
This expression incorporates the same Euclidean two-point function
except that the singularities now arise as the sources
$\tphi_{0\pm}(\Omega)$ approach antipodes on the $n$-sphere.

These results suggest that one need only consider a single copy of
the CFT and that an operator on $I^+$ is identified with the same
operator on $I^-$ after an antipodal mapping. One finds further
support for this interpretation by considering the isometries of
dS space. For example, the isometry\footnote{This isometry
corresponds to the action of a time translation $\partial_t$ in
the static patch coordinates \cite{AS5}.} which produces a
dilatation around a point on $I^-$. On $I^+$, the same symmetry
corresponds to a dilatation around the antipodal point on the
$n$-sphere.

However, this suggestion for identifying operators at $I^+$ and $I^-$
is easily seen to require some
revision as follows. As discussed in the introduction, bulk
correlators are naturally related by time evolution. The key
ingredient is simply the free field evolution of the scalar, which
given some configuration specified on a $n$-dimensional
hypersurface is characterized by the formula
\beq \labell{evolforward}
\phi(x') = \int d\Sigma^{\mu}\
\phi(x)\stackrel{\leftrightarrow}{\partial}_{\mu}G_{R}(x,x')\ ,
\eeq
where $G_{R}(x,x')$ is a retarded Green's function, \ie it
vanishes for $t>t'$. Now as an example, the integral appearing in
the generating functional \reef{evolmap} is covariant and so
should be invariant when evaluated on any time slices $t$ and
$t'$. The advantage of pushing these slices to $I^-$ (or $I^+$)
lies in the fact that one can easily separate the source currents
according to their conformal weights.

We can explicitly consider the relation between currents on the
past and future boundaries by simply following the classical
evolution \reef{evolforward} of the fields from $I^-$ to $I^+$.
Unfortunately, it is clear that generically there is no simple
local relation between the currents on $I^-$ and those on $I^+$.
This remark comes from the observation that in general the
retarded Green's function will have support throughout the
interior of the light cone. This intuition is readily confirmed by
explicit calculations. Ref.~\cite{tagirov} presents explicit
Green's functions for generic masses in four-dimensional de Sitter
space. So, for example, for scalar fields in the principal series,
the retarded Green's function becomes for large timelike proper
separation
\beq \labell{green22}
G_{R}(t,\Omega;t',\Omega') \propto
\frac{\sin^{n/2} \tau\,\sin^{n/2}\tau'} {(w^{i}w^{\prime i}-\cos
\tau\,\cos \tau')^{n/2}}\,\theta(\tau'-\tau)
\ , \eeq
where $\tau$ is the conformal time coordinate, $\sin\tau=1/\cosh
t/\lc$ --- see eq.~\reef{talespace}, below. Here the
$\theta$-function ensures the proper time-ordering of the points.
In any event, eq.~\reef{green22} illustrates how the field `leaks'
into the interior of the lightcone with the classical evolution.
Generically this leads to a nonlocal mapping between the currents
on $I^-$ and $I^+$. This complication will only be avoided in
certain exceptional cases, for example, if the retarded Green's
function has only support precisely on the light cone
--- a point to which we return below.

The nonlocal relation between the currents on $I^-$ and those on
$I^+$ can be made more explicit through the mode expansion of the
fields on dS space --- see appendix \ref{appendix1}. A
well-documented feature of cosmological spacetimes is mode-mixing
or particle creation \cite{BD}. For the present case of dS space,
this corresponds to the fact that a mode of the scalar field with
a given boundary behavior on $I^{-}$, \eg having $h_{-}$ scaling,
will usually have a mixture of $h_\pm$ scaling components at
$I^+$. Appendix~\ref{appendix1} provides a detailed discussion of
the mode expansions on dS space as well as the Bogolubov
transformation relating the modes with a simple time dependence
(scaling behavior) near $I^-$ to those near $I^+$. Using these
results, we may discuss the mapping between the currents on the
conformal boundaries. Following the notation of
appendix~\ref{appendix1}, we decompose the asymptotic fields in
terms of spherical harmonics on the $n$-sphere
\beq \labell{decomp1}
\phi_{0\pm}(\Omega)=\sum_{L,j} a_{\pm Lj}
Y_{Lj} \ ,\qquad\qquad \tphi_{0\pm}(\Omega)=\sum_{L,j}
\tilde{a}_{\pm Lj} Y_{Lj}\ .
\eeq
Denoting the antipodal map on the $n$-sphere as $\Omega\rightarrow
J\Omega$, one has\footnote{This result becomes clear when the
$n$-sphere is embedded in $R^{n+1}$ with
$(x^1)^2+(x^2)^2+\cdots+(x^{n+1})^2=1$. In this case, the
spherical harmonics $Y_{Lj}$ may be represented in terms of
symmetric traceless tensors, $Z_{i_1i_2\cdots i_L}
x^{i_1}x^{i_2}\cdots x^{i_L}$, and hence it is clear that the
antipodal map, which takes the form $J:x^i\rightarrow -x^i$,
produces an overall factor of $(-)^L$.}
$Y_{Lj}(\Omega)=(-)^L\,Y_{Lj}(J\Omega)$. Now let us imagine that
$\phi_{0\pm}$ and $\tphi_{0\pm}$ are related by the antipodal map,
\ie $\phi_{0\pm}(\Omega)=z\,\tphi_{0\pm}(J\Omega)$ with some
constant phase $z$. Then one must have
\beq\labell{antip}
{a}_{\pm Lj}=z\,(-)^L\,\tilde{a}_{\pm Lj}
\eeq
where in particular the constant $z$ is independent of $L$.

However, in general, the Bogolubov transformation given in
appendix~\ref{appendix1} gives a more complex mapping. For
example, from eq.~\reef{bogprinc} for the principal series, one
finds
\beq \labell{modemix11}
{a}_{\pm Lj}= C_{-}^{-}(\omega) e^{\pm2i\theta_L}\tilde{a}_{\pm Lj}
+C_{-}^{+}(\omega) \tilde{a}_{\mp Lj}\ .
\eeq
Now given eq.~\reef{deven} for $n$ odd with both
$C_{-}^{-}(\omega)$ and $C_{-}^{+}(\omega)$ nonvanishing,
certainly eq.~\reef{antip} is inapplicable. One comes closer to
realizing the desired result with even $n$ for which
$C_{-}^{-}(\omega)=1$ and $C_{-}^{+}(\omega)=0$. However, for
either $n$ odd or even, the phase $\theta_L$ always introduces a
nontrivial $L$ dependence (beyond the desired $(-)^L$) as shown in
eq.~\reef{phase}. Thus while the mapping between $I^-$ and $I^+$
may look relatively simple in this mode expansion, it will clearly
be nonlocal when expressed in terms of the boundary data
$\phi_{0\pm}(\Omega)$ and $\tphi_{0\pm}(\Omega)$.

The complementary series gives some more interesting
possibilities with
\beqa
a_{-Lj} &=& \bar{C}_{-}^{-}(\mu)\, \tilde{a}_{-Lj} +
\bar{C}_{-}^{+}(\mu)\, \tilde{a}_{+Lj}\ ,
\nonumber\\
\labell{modemix22} a_{+Lj} &=& \bar{C}_{+}^{-}(\mu)\,
\tilde{a}_{-Lj} + \bar{C}_{+}^{+}(\mu)\, \tilde{a}_{+Lj}\ .
\eeqa
In particular for $n$ odd and $\mu$ half integer, one finds
$\bar{C}_{-}^{+}(\mu)=0=\bar{C}_{+}^{-}(\mu)$ and
$\bar{C}_{-}^{-}(\mu)=(-)^{{n\over2}+\mu}(-)^L=\bar{C}_{+}^{+}(\mu)$.
Note that these special cases include $\mu=1/2$, which corresponds
to the conformally coupled massless scalar field to which we will
return in the following section. Similarly for $n$ even and $\mu$
integer:
$\bar{C}_{-}^{-}=\bar{C}_{+}^{+}=(-1)^{{n\over2}+\mu+1}(-1)^L$ and
$\bar{C}_{-}^{+}=0=\bar{C}_{+}^{-}$. Hence the coefficients for
these special cases give a precise realization of
eq.~\reef{antip}. Further for these cases then, the generating
functional considered in eq.~\reef{evolmap2} will take the simple
form given in eq.~\reef{mapans99}.

Hence when considering the principal series or generic masses in
the complementary series, it seems that nonlocality will be an
unavoidable aspect of the relation between field theory operators
associated with two conformal boundaries. The essential point is
that the time evolution of the scalar generically introduces
nonlocality in the mapping because the retarded Green's function
smears a point-like source on $I^-$ out over a finite region on
$I^+$. However, note that one reproduces precisely the same
boundary correlators but after some nonlocal reorganization of the
degrees of freedom within the dual field theory. It seems
appropriate to refer to such relations as {\it nonlocal dualities}
within the field theory. On the other hand, the complementary
series does seem to provide some situations where the mapping of
the boundary data between $I^-$ and $I^+$ is local. In the absence
of a working example of the proposed dS/CFT duality, one might
interpret these results as a hint towards the specific types of
fields that would appear in a successful realization of the
dS/CFT. Unfortunately, however, this selection rule based on
locality of the mapping between boundaries does not seem to
survive in more interesting applications, as we will see in the
following.

\subsection{Nonlocal dualities in `tall' spacetimes}

It is of interest to extend the application of the dS/CFT
correspondence from dS space to more general spacetimes with
asymptotically dS regions. As a consequence of a theorem of Gao
and Wald \cite{GW}, such a (nonsingular) background will be `tall'
\cite{tale1}. That is, the conformal diagram for such spacetimes
must be taller in the timelike direction than it is wide in the
spacelike direction. Of course, this feature has important
implications for the causal connection between the past and future
boundaries, and hence for the relation between the dual field
theory operators defined at these surfaces. In particular, the
latter relation becomes manifestly nonlocal.

We may explicitly illustrate the causal structure of the tall
spacetimes by working in conformal coordinates. For asymptotically
dS spacetimes which are homogeneous on spherical hypersurfaces,
the metric may be written
\beq \labell{talespace}
ds^{2} = C(\tau)\left[ -d\tau^{2}+ d\theta^2+\sin^2\theta\,
d\Omega^{2}_{n-1}\right]\ .
\eeq
For pure dS space, $C(\tau)=\lc^2/\sin^2\tau$. Note that the
proper time $t$ in eq.~\reef{metricds} is related to the conformal
time $\tau$ above by the coordinate transformation
$\sin\tau=1/\cosh t/\lc$.

In this case, the conformal time runs from $\tau=0$ at $I^-$ to
$\tau=\pi$ at $I^+$. The angular coordinate $\theta$ on the
$n$-sphere runs over the same range, \ie from $\theta=0$ at the
north pole to $\theta=\pi$ at the south pole. Hence it is clear
that the conformal diagram for dS space is a square (see figure
\ref{heights}).

Now for a tall spacetime, the conformal time above would run over
an extended range $0\leq\tau\leq\pi+\Delta$ where $\Delta>0$. The
assumption that the background is asymptotically dS means that the
conformal factor has the following behavior near $I^\pm$:
\beqa
\labell{bcasds}
\lim_{\tau\rightarrow 0} C(t) &=&
\frac{\lc^2}{\sin^{2}\tau}\ ,\\
\lim_{\tau\rightarrow
\pi + \Delta} C(t) &=& \frac{\tlc^2}{\sin^{2} (\tau-\Delta)}\ ,
\nonumber
\eeqa
where we have allowed for the possibility that the cosmological
`constant' is different at $I^+$ than at $I^-$. This possibility
may be realized in a model where a scalar field rolls from one
critical point of its potential to another \cite{tale1}. In any
event, the corresponding conformal diagram will be a rectangle
with height $\delta\tau=\pi+\Delta$ and width $\delta\theta=\pi$
(see figure \ref{heights2}).

This increase in the height of the conformal diagram modifies the
causal connection between $I^\pm$ in an essential way. Consider
the null rays emerging from the north pole ($\theta=0$) at $I^-$
($\tau=0$). This null cone expands out across the $n$-sphere
reaching the equator ($\theta=\pi/2$) at $\tau=\pi/2$, and then
begins to reconverge as it passes into the southern hemisphere.
The null rays focus at the south pole ($\theta=\pi$) at
$\tau=\pi$, however, in this tall spacetime, this event
corresponds to a finite proper time for an observer at the south
pole. Beyond this point, the null cone expands again and
intersects $I^+$ ($\tau=\pi+\Delta$) on the finite-sized
($n$--1)-sphere at $\theta=\pi-\Delta$.
\FIGURE{\epsfig{file=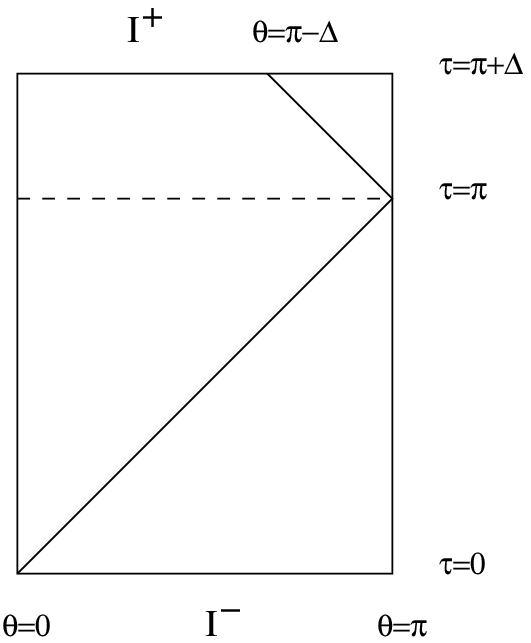, width = 5cm}\caption{Conformal
diagrams of a perturbed de Sitter space. The excess height is
represented by $\Delta$.} \label{heights2}}

The discussion of the previous section made clear that an
essential ingredient in finding a simple local mapping of boundary
data on $I^-$ to that on $I^+$ in dS space was the refocusing of
the above null cone precisely at the future boundary. Even in that
case, we pointed out that the time evolution of the scalar
generically introduces nonlocality in the mapping because the
retarded Green's function smears a point-like source on $I^-$ out
over a finite region on $I^+$. Here we see that in a tall
spacetime, a nonlocal map is inevitable since the causal
connection between the past and future boundaries is itself
nonlocal. So we should expect that even in the special cases found
to have a local map for pure dS space, the mapping should become
nonlocal for these same theories in a tall background. That is,
for these more general asymptotically dS spacetimes, the relation
between the dual field theory operators defined at each of the
boundaries becomes nonlocal. Hence we are naturally lead to
consider a nonlocal self-duality of the CFT. Further we note that
given the results of Gao and Wald \cite{GW} this would be the
generic situation. For example, injecting a single scalar field
quantum into dS space would actually lead to backreaction effects
which would produce a tall spacetime.

\subsection{The conformally coupled massless scalar}

We now turn to consider conformally coupled massless scalar field
theory as an example which illustrates several of the points
discussed above. In particular, it is an example where the mapping
between the past and future boundaries is local in pure dS space,
but becomes nonlocal in a tall background. Another useful feature
is that one can perform explicit calculations in a tall spacetime
without referring to the detailed evolution of the conformal
factor $C(\tau)$. Rather a knowledge of the boundary conditions
\reef{bcasds} is sufficient.

The conformally coupled massless scalar corresponds to the
curvature coupling $\xi = \frac{n-1}{4n}$ and $m^2=0$ in
eq.~\reef{messs}. Hence in pure dS space or in an asymptotically
dS region, $M^2\lc^2 = (n^2-1)/4$ and the corresponding scaling
exponents \reef{exponents} become $h_\pm=(n\pm 1)/2$, independent
of the value of the cosmological constant. As one might infer from
the real exponents, this field lies in the complementary series
for any value of the cosmological constant. The remarkable
property of this scalar field theory is that the solutions of the
wave equation \reef{eomfirst2} transform in a simple way under
local conformal scalings of the background metric \cite{BD}.

The backgrounds of interest \reef{talespace} are conformally
flat\footnote{Note that the coordinate transformation $T=e^\tau$
puts the metric \reef{talespace} in the form of the flat Milne
universe, up to a conformal factor.} and therefore the Green's
function describing the evolution in the tall background is simply
the flat space Green's function for a {\it massless} scalar field,
up to some overall time dependent factors. In particular then, for
$d$ even ($n$ odd), the Green's function will have support
precisely on the light cone. For example, in four-dimensional dS
space, the retarded Green's function can be written as
\beq \labell{mccgreen2} G_{R}(\tau,\Omega;\tau',\Omega') =
-\frac{\sin \tau \sin \tau'}{4\pi\lc^2} \,  \delta(w^{i}w^{\prime
i} - \cos (\tau'-\tau))\,\theta(\tau'-\tau)\ . \eeq
Similarly in higher even-dimensional dS spaces, the Green's
function will contain $\delta$-functions (and derivatives of
$\delta$-functions) with support only on the light cone
\cite{junkkk}. Given this form of the retarded Green's functions,
the evolution of the scalar field \reef{evolforward} from $I^-$ to
$I^+$ will produce precisely the antipodal mapping for all of
these cases. Note that this result is confirmed by the mode
analysis in the first part of this section. The conformally
coupled massless scalar has $\mu=1/2$ and we are considering even
dimensions or $n$ odd. This combination matches one of the special
cases in which the modes transformed according to the antipodal
mapping.

Using the conformal transformation properties of the field
\cite{BD}, the analogous Green's function for any spacetime of the
form \reef{talespace} is easily constructed. For $d=4$, it may be
written as
\beq \labell{gengreenmcc} G_{R}(\tau,\Omega;\tau',\Omega') =
-\frac{1}{4\pi}  \frac{1}{\sqrt{C(\tau)}\sqrt{C(\tau')}}
\,\delta(w^{i}w^{\prime i} - \cos
(\tau'-\tau))\,\theta(\tau'-\tau)\ . \eeq
For other values of even $d$, the corresponding Green's function
has a similar form. For the conformally coupled scalar in such
tall spaces, the delocalization of the boundary map does not
depend on the detailed evolution, \ie the details of $C(\tau)$.
Rather the nonlocality is completely characterized by $\Delta$,
the excess in the range of the conformal time. For example, a
source current placed at the north pole ($\theta=0$) on $I^-$ is
smeared over an ($n$--1)-sphere centered at the south pole
($\theta=\pi$) and of angular radius $\delta\theta=\Delta$ on
$I^+$.

Using eq.~\reef{gengreenmcc}, we can make this discussion
completely explicit for four dimensions. Consider an arbitrary
tall space \reef{talespace} with $n=3$ satisfying the boundary
conditions given in eq.~\reef{bcasds}. First with the conformal
time coordinate, the asymptotic boundary conditions
\reef{boundary-} for the scalar field at $I^-$ become
\beq \labell{newbou-} \lim_{\tau \rightarrow 0}\phi(\Omega,\tau)
\simeq \phi_{0+}(\Omega)\,\tau^{h_+}+\phi_{0-}(\Omega)
\,\tau^{h_{-}} \eeq
and similarly at $I^+$, we have
\beq \labell{newbou+} \lim_{\tau \rightarrow
\pi+\Delta}\phi(\Omega,\tau) \simeq \tilde\phi_{0+}(\Omega)
\,(\pi+\Delta-\tau)^{h_+}+\tilde\phi_{0-}(\Omega)
\,(\pi+\Delta-\tau)^{h_{-}}\ . \eeq
These boundary conditions apply for a general scalar field theory.
In the present case of a conformally coupled massless scalar with
$n=3$, we have $h_+=2$ and $h_-=1$.  Hence inserting
(\ref{gengreenmcc}) and (\ref{newbou-}) into (\ref{evolforward}),
we may evaluate the result at a point $(\Omega', \tau' = \pi +
\Delta - \epsilon)$ near $I^+$ and compare to eq.~\reef{newbou+}.
The final result for the boundary fields on $I^+$ is
\begin{eqnarray}
\tilde\phi_{0+}(\Omega')&=&\frac{|\sin\Delta|}{\sin\Delta}
\frac{\ell}{\tilde \ell}
\bigg\{
\sin \Delta \, \langle\phi_{0-} \rangle_\Delta (J \Omega' )
- \cos \Delta \, \langle\phi_{0+} \rangle_\Delta (J\Omega')
\bigg\}
\labell{hungry}\\
\tilde\phi_{0-}(\Omega')&=&\frac{|\sin\Delta|}{\sin\Delta}
\frac{\ell}{\tilde \ell}\bigg\{
\cos\Delta \, \langle\phi_{0-} \rangle_\Delta (J \Omega')
+ \sin\Delta \, \langle\phi_{0+} \rangle_\Delta (J\Omega')
+\sin\Delta \, \partial_{\theta} \langle \phi_{0-} \rangle_\Delta (J\Omega')
\bigg\}\ ,
\nonumber
\end{eqnarray}
where $J$ is the antipodal map on the two-sphere and $\langle
\phi_{0\pm} \rangle_\Delta(J \Omega')$ denote the average of
$\phi_{0\pm}$ on the two-sphere separated from $J \Omega'$ by an
angle  $\Delta$.
The factors $\frac{ |\sin \Delta|}{\sin \Delta}$ are to be
understood as being continuous from below; \ie this factor is $-1$
at $\Delta =0$ and $+1$ at $\Delta = \pi$.

This expression simplifies tremendously in the case of
dS space with $\Delta =0$ (as well as $\tilde\ell=\ell$) to
yield
\begin{equation}
\tilde\phi_{0+}(\Omega')=\phi_{0+}(J\Omega')
\ ,\qquad\qquad
\tilde\phi_{0-}(\Omega')=-\phi_{0-}(J\Omega')
\ .
\labell{simplea}
\end{equation}
Thus, in pure four-dimensional dS space the map from $I^-$ to $I^+$ acts
on the conformally coupled massless scalar field as simply
the antipodal map on $\phi_{0+}$ and $-1$ times the antipodal map
on $\phi_{0-}$. Note that the time reflection symmetry of de Sitter allows
solutions for the mode functions to be decomposed into even and odd parts and,
furthermore, both even and odd solutions will exist.
Thus, with our conventions and $h_\pm$ real,
when evolution from $I^-$ to $I^+$
leads to the antipodal map it will be associated with a phase $z = +1$ for
one set of modes and the opposite phase $z=-1$ for the other.

\section{Discussion}
\label{disc}

The dS/CFT correspondence is a striking proposal which carries the
potential for extraordinary new insights into cosmology and the
cosmological constant problem. Unfortunately, the outstanding
problem remains to find a concrete example where the bulk gravity
theory and the dual field theory are understood or at least known.
Lacking the guidance that such a working model would provide, one
is left to study various aspects of physics in (asymptotically) dS
spacetimes from this new point of view and to determine properties
which this correspondence implies for the dual Euclidean CFT.

Such investigations have yielded a number of unusual properties
for the dual field theory. It is likely to be nonunitary, \eg if
the bulk theory involves scalars in the principal series
\cite{AS1}. A nonstandard inner product is required to reproduce
ordinary quantum field theory in the bulk \cite{AS3,annals}. One
might also observe that this Euclidean field theory should not
simply be a standard Wick rotation of a conventional field theory
since attempting to `un-Wick rotate' would produce a bulk theory
with two time directions and all of the associated confusions. We
may add to this list the observation of section \ref{dSrev} that,
since bulk correlators are not symmetric in Lorentz signature
quantum field theory, a straightforward duality would require
non-symmetric correlation functions in the dual Euclidean theory.
But correlators generated by functional differentiation of a
partition function are always symmetric, so the Euclidean theory
could have no definition through a partition sum. Finally, in the
present paper, we have also inferred the existence of unusual
nonlocal dualities within the field theory itself.

Our investigation focussed on the mapping of operators between
$I^+$ and $I^-$ provided by time evolution in the bulk spacetime.
The essential point is that the time evolution of the scalar
generically introduces nonlocality in the mapping because the
retarded Green's function smears a point-like source on $I^-$ out
over a finite region on $I^+$. However, despite this nonlocal
reorganization of the degrees of freedom within the dual field
theory, one reproduces the same boundary correlators. Hence we
referred to this relation as a {nonlocal duality} within the field
theory. While this nonlocality already applies for many fields in
pure dS space, it seems unavoidable in tall spacetimes because the
causal connection between $I^+$ and $I^-$ is inherently nonlocal.
We emphasize that tall spacetimes are quite generic as a result of
the theorem in \cite{GW}. As soon as one perturbs dS even slightly
by, \eg the introduction of matter fields or gravitational waves,
the resulting background solution will have the property that its
conformal diagram is taller than it is wide. As the inferred
self-duality is nonlocal, \ie local operators are mapped to
nonlocal operators, it seems that the underlying field theory does
not have a unique concept of locality. That is, one has a specific
dictionary whereby the same short distance singularities can be
reproduced by a set of local or nonlocal operators.

Faced with the daunting task of consolidating all of these unusual
characteristics in a single Euclidean field theory, one is tempted
to revise the interpretation of the dS/CFT correspondence. One
suggestion \cite{annals} is that the duality should involve two
CFT's but that dS spacetime is defined as a correlated state in
Hilbert space of the two field theories. The correlated state is
constructed so as to preserve a single SO($n+1,1$) symmetry group,
which is then reflected in the isometries of the dS space. As
discussed in section \ref{nonlocal}, we still feel that our
experience with the AdS/CFT is highly suggestive that the two
boundaries should not be associated with distinct field theories.
Further, it is difficult to see how this framework could
incorporate big bang or big crunch backgrounds with a single
asymptotic dS region. Note that the latter spacetimes will still
give rise to horizons, as well as the associated thermal radiation
and entropy.

However, this approach with two CFT's remains an intriguing
suggestion. Within this context, the mapping of the boundary data
between $I^\pm$ would provide information about correlations in
the field theory state. Hence our calculations would still find
application in this context. The nonlocalities discussed here,
while not unnatural, give an indication of the complexity of
these correlations.

We should also remark that all of our investigations treated only
the time evolution of a free scalar field theory. The mapping of
boundary operators will become even more complex if one was to
consider an interacting field theory. Of course, in accord with
the discussion here, we would still expect that time evolution of
the fields or operators in an interacting theory would still
provide the basis for this mapping.

While it is amusing to speculate on such matters, we note that the
central thesis of \cite{obstruct} is that one cannot successfully
understand the physics of dS space within the context of quantum
field theory in curved spacetime. It is interesting to consider
how their comments may relate our discussion. Essentially, they
suggest that bulk properties of dS space should be analogous to
the physics in a thermal system with a finite number of states and
deduce from this that the evolution map of linearized quantum
field theory should not be trusted in detail near the past and
future boundaries.  As a result, they suggest that a dual theory
may not be as local as one might expect by studying limits of bulk
correlation functions in background quantum field theory. The
comments of \cite{fantastic} raise further questions about
correlation functions between points with a large separation in
time.  In particular, the problematic correlators would include
precisely those between operators on $I^-$ and $I^+$. Here the
smearing observed in the tall spacetimes is likely to play a role
since, if backreaction is properly accounted for, even injecting a
single scalar field quantum into dS should deform it to a
(slightly) tall spacetime. It may be that the nonlocalities
discussed here may be a hint that the `correct physical
observables' are themselves nonlocal\footnote{Similar implications
can be drawn from the finite time resolution discussed in
\cite{nightmare}.} so that the boundary map would preserve the
form of such operators.

Note that there is a certain tension between our strong reliance
on time evolution, through which observables near any two Cauchy
surfaces can be related, and the idea that the bulk evolution is
related to a renormalization group flow in the dual theory
\cite{AS,BBM}. The point is that time evolution naturally produces
a scaling of distances on Cauchy surfaces (at least in simple
examples) and so these surfaces are naturally associated with
different distance scales in the dual theory. However, the time
evolution map relating different surfaces is invertible. In
contrast, the usual notion of the renormalization group is
actually that of a semi-group, in which different scales are
related by integrating out modes, \ie by throwing away short
distance details so that the descriptions at two different scales
are not fully equivalent.

To gain some perspective on this issue, we would like to return
briefly to the AdS/CFT case and the interpretation of
renormalization group flows.  Recall that the primary assumption
is that the relevant asymptotically AdS spacetime is in fact dual
to the vacuum of some field theory.  The important point is that
one begins by placing the entire spacetime in correspondence with
the vacuum of some single theory.  One then uses the IR/UV
connection to argue that different regions of the bulk spacetime
are naturally related to different energy regimes in the dual
theory.   The suggestion that this description at differing energy
scales is somehow connected to a renormalization group flow seems
natural and, in that context, there was no evolution map relating
the inner and outer regions to provide such an obvious tension.

In the present dS/CFT context, such a tension does exist. However,
the more primitive association of different parts of the spacetime
with the behavior of the field theory at differing energy scales
still seems plausible.  A more concrete version of this idea is
suggested by the behavior of the bulk evolution map itself.  As we
have seen, the evolution map from $t$ to $t'$ `coarse grains' the
observables on $t'$ in the sense that the theory is now presented
in terms of variables (those that are local at time $t$) which are
nonlocal averages over the intersection of past light cones from
time $t$ with the original hypersurface at $t'$.  However, a
sufficient number of overlapping coarse grainings are considered
that no information is lost. Such a procedure can also be
performed in a Euclidean field theory and one might speculate that
keeping only the simplest terms in the resulting action might bare
some similarity to those obtained from more traditional
renormalization group methods. This would be in keeping with the
identification of a c-function \cite{AS,BBM,tale1} in which a
spacetime region was associated with a copy of de Sitter space by
considering only the metric and extrinsic curvature on a
hypersurface.

Note that this interpretation readily allows us to run our flow
both `forward' (toward the IR) and `backward' (toward the UV).
However, it is far from clear that the coarse graining procedure
is unique. This fits well with the interpretation suggested in
\cite{tale1} for `renormalization group flow spacetimes' with
spherical homogeneity surfaces. There, one naturally considers two
UV regions (one at $I^-$ and one at $I^+$) which both `flow' to
the same theory at some minimal sphere where the two parts of the
spacetime join.  One simply reads the flow as starting in the UV,
proceeding toward the IR, but then reversing course.
Interestingly, it is possible to arrive at a different UV theory
from which one began.  Such an odd state of affairs is more
natural when one recalls that we have already argued that the
theory must possess a nonlocal duality, so that it in fact has two
distinct local descriptions.

\acknowledgments

The authors would like to thank Vijay Balasubramanian, Cliff
Burgess, Raphael Bousso, Oliver de Wolfe, Laurent Freidel, Ted
Jacobson, Lev Kofman, Gilad Lifschytz, Alex Maloney, Eric Poisson,
Phillipe Pouliot, Simon Ross and Andy Strominger for interesting
conversations. We also thank David Winters for help with the
figures. FL and RCM were supported in part by NSERC of Canada and
Fonds FCAR du Qu\'ebec. DM was supported in part by NSF grant
PHY00-98747 and funds from Syracuse University. DM would like to
express his thanks to the Perimeter Institute for their warm
hospitality at the initial and final stages of this work. FL would
like to thank the University of Waterloo's Department of Physics
for their ongoing hospitality during this project. Finally, all of
the authors would like to thank the Centro de Estudios Cient\'\i
ficos in Valdivia, Chile for their hospitality at an intermediate
stage of this work.

\appendix

\section{Scalar field modes in dS space}
\label{appendix1}

In this appendix, we present a detailed analysis of the bulk
physics of massive scalar fields propagating in a dS space of
arbitrary dimension, emphasizing characteristics of their
evolution which should be relevant to the proposed dS/CFT
correspondence. Our aim is to characterize fully the mode mixing
phenomenon inherent to physics in dS space. While the details of
this analysis are readily available in the literature for the
modes of the principal series (see, for example, ref.~\cite{AS3}),
we did not find explicit accounts of the complementary and
discrete series.

\subsection{Field equation and asymptotic behavior}

The spherical foliation of ($n$+1)-dimensional dS space is given by
the metric
\beq
\labell{ametrica}
ds^{2} = -dt^{2} +\cosh^{2}t\ d\Omega^{2}_{n},
\eeq
where in this appendix we set the dS radius to unity ($\lc =1$).
We consider a massive scalar field propagating in this background
according to
\beq
\labell{eomfirst}
\left[ \Box - M^{2} \right] \phi(x) = 0.
\eeq
It is convenient to write the solutions to eq.~(\ref{eomfirst}) in the form
\beq
\labell{modality}
\phi(x) = y_{L}(t)\ Y_{Lj}(\Omega),
\eeq
where the $Y_{Lj}$'s are spherical harmonics on the $n$-sphere satisfying
\beq
\labell{lap}
\nabla^{2} Y_{Lj} = -L(L+n-1) Y_{Lj},
\eeq
where $\nabla^2$ is the standard Laplacian on the $n$-sphere.
The differential equation for $y_{L}(t)$ is then
\beq
\labell{dems}
\ddot{y}_{L} + n\tanh t \; \dot{y}_{L} + \left[ M^{2} +
\frac{L(L+n-1)}{\cosh^{2}t} \right]y_{L} = 0.
\eeq
As discussed in section 2.2,
of particular relevance to the dS/CFT correspondence is the behavior of the scalar
field near the boundaries $I^{+}$ and $I^{-}$ as $t\rightarrow \pm \infty$.
In these limits, eq.~(\ref{dems}) becomes
\beq
\labell{asympeom}
\ddot{y}_{L} \pm n\dot{y}_{L} + M^{2} y_{L} = 0,
\eeq
which implies that
\beq
\labell{simpletime}
\lim_{t\rightarrow -\infty} y_{L} \sim e^{h_{\pm}t}, \;\;\;\;\;\;\;
\lim_{t \rightarrow +\infty} y_{L} \sim e^{-h_{\pm}t},
\eeq
where the weights $h_{\pm}$ are defined by
\beq
\labell{water}
h_{\pm} = \frac{n}{2} \pm \sqrt{\frac{n^{2}}{4} - M^{2}}\equiv\frac{n}{2} \pm \mu.
\eeq

Formally, one may classify scalar fields according to the
irreducible representations of SO($n+1,1$), the isometry group of
de Sitter space, which are labelled by the eigenvalues associated
with the Casimir operator\footnote{In fact, there are two
coordinate invariant Casimir operators associated with the de
Sitter isometry group but only one is relevant in characterizing
massive scalar fields. The other Casimir operator automatically
vanishes for all spin-zero fields but may play a role in the
classification of higher spin representations \cite{gazeau1}.
Another interesting formal question is the behavior of these
representations in the limit where the cosmological constant is
taken to zero. A complete treatment of representation contraction
in de Sitter space can be found in ref.~\cite{MN}.} $Q=\Box$,
which simply corresponds to the mass parameter $M^{2}$. The {\it
principal series} is defined by the inequality $M^{2} > n^{2}/4$.
In this case, the weights $h_{\pm}$ have an imaginary part, and
the corresponding modes, while still being damped near the
boundaries, have an oscillatory behavior in the bulk. For the {\it
complementary series}, the effective mass falls in the range $0 <
M^{2} \leq n^2/4$. As will be made more explicit later, the modes
are non-oscillatory asymptotically in this case since both
$h_{\pm}$ are real quantities. The remaining {\it discrete series}
corresponds to $M^{2}<0$. This last condition means that $h_{-}<0$
(and $h_{+}>n$), which implies that the tachyonic fields scaling
like $y_{L} \sim e^{\mp h_{-}t}$ in approaching on $I^{\pm}$ are
growing without bound. Still, one is able to find `normalizable'
modes in a certain limited number of cases, as will be discussed
below.

The case of $M^2=0$, \ie a massless scalar field, is interesting and deserves
further comment. One finds that in this case it is impossible to construct a vacuum
state which is invariant under the full de Sitter group SO($n+1$,1). A great deal of
discussion about the peculiar nature of this quantum field theory can be found in
the literature \cite{GRT,turok,allen2}. The weights associated with the $M^{2}=0$ field
are $h_{+}=n$ and $h_{-}=0$. Dual to the latter should be a
marginal operator in the CFT, \ie a deformation which does not scale under conformal
transformations.

To fully solve eq.~\reef{dems}, we make the change of variables
\beq
\labell{full}
y_{L}(t) = \cosh^{L}\!t\,  e^{(L+\frac{n}{2}+ \mu)t}g_{L}(t).
\eeq
Setting $\sigma=-e^{2t}$, this equation for the time dependant profile
takes the form of the hypergeometric equation:
\beq
\labell{diffhyper}
\sigma(1-\sigma) g'' + \left[ c - (1+a+b) \sigma \right] g' - ab g = 0,
\eeq
where a `prime' denotes a derivative with respect to $\sigma$ and the coefficients are
\beq
a = L + \frac{n}{2}, \;\;\;\;\; b = L + \frac{n}{2} + \mu, \;\;\;\;\; c = 1 + \mu.
\eeq
The two independent solutions can then be expressed
in terms of hypergeometric functions,
\beq
\labell{sol11}
y_{L+}(t) = N_{+} \cosh^L\!t\, e^{(L+h_{+})t} F(L+\frac{n}{2},L+h_{+};1+\mu;-e^{2t}),
\eeq
\beq
\labell{sol12}
y_{L-}(t) = N_{-} \cosh^L\!t\, e^{(L+h_{-})t} F(L+\frac{n}{2},L+h_{-};1-\mu;-e^{2t}),
\eeq
where $N_{\pm}$ are normalization constants, which will be fixed below.
More specifically, we have chosen here
the two linearly independent solutions of eq.~(\ref{diffhyper}) in the neighborhood
of the singular point $-e^{2t}= 0$ \cite{handbook}, which corresponds to one of the
two limits of interest, \ie  $t \rightarrow - \infty$.
Following eq.~\reef{modality}, we denote the complete mode functions as $\phi_{L\pm} = y_{L\pm}(t)
Y_{Lj}(\Omega)$.

One important aspect of time evolution of the scalar field in the
bulk is the mode mixing that occurs between the two boundaries,
$I^\pm$. For example, this would be related to the particle
production in the dS space \cite{AS3,allen2,em1,allen}. In the
following, we emphasize the differences between the principal,
complementary and discrete series.

\subsection{Principal series}

The principal series is frequently discussed in the physics literature, \eg
\cite{AS3,allen2,em1,allen}, and would seem to be the most relevant case for the particle
spectrum observed in nature. We review some of the salient points here for comparison with
the other representations in the following
subsection. For the principal series, it is useful to introduce $\omega\equiv-i\mu$.
Then the above modes become
\beq
y_{L-}(t) = \frac{2^{L+(n-1)/2}}{\sqrt{\omega}} \cosh^{L}\!t\, e^{(L+\frac{n}{2} - i\omega)t}
F(L+\frac{n}{2},L+\frac{n}{2}-i\omega;1-i\omega;-e^{2t}),
\eeq
\beq
y_{L+}(t) = \frac{2^{L+(n-1)/2}}{\sqrt{\omega}} \cosh^{L}\!t\, e^{(L+\frac{n}{2} + i\omega)t}
F(L+\frac{n}{2},L+\frac{n}{2}+i\omega;1+i\omega;-e^{2t}),
\eeq
where $y_{L-}^{*}(t) = y_{L+}(t)$. Here the normalization
constants have been fixed by imposing
$(\phi_{L+},\phi_{L+})=1=(\phi_{L-},\phi_{L-})$ as usual with the
standard Klein-Gordon inner product \cite{BD}. As emphasized
above, these solutions have the simple time dependence of
eq.~\reef{simpletime} in the asymptotic region
$t\rightarrow-\infty$ near $I^-$. Because the differential
equation (\ref{dems}) is invariant under $t\rightarrow -t$, one
can easily define another pair of linearly independent solutions
by applying this transformation to the above modes. We label the
resulting modes: $y_{L}{}^{-}(t)$ and
$y_{L}{}^{+}(t)=y_{L}{}^{-*}(t)$,  where
\beq
y^{-}(t) = y_{+}^{*}(-t).
\eeq
It readily follows that $y_{L}{}^{-}\sim e^{-h_{-}t}$ and
$y_{L}{}^{+}\sim e^{-h_{+}t}$ near $I^{+}$. The two sets of modes
$y_{L\pm}(t)$ and $y_{L}{}^{\pm}(t)$ can respectively be used to
construct the `in' and `out' vacua with no incoming and outgoing
particles. The Bogolubov coefficients relating these two sets of
modes are defined through
\beq
\labell{bogprinc}
y_{L-}(t) = C_{-}^{-}(\omega)\, e^{-2i\theta_{L}}\,y_{L}{}^{-}(t) + C_{-}^{+}(\omega)\,
y_{L}{}^{+}(t),
\eeq
with a similar expression for $y_{L+}$ (with
$C_{+}^{+}(\omega)=C_{-}^{-}(\omega)$ and
$C_{+}^{-}(\omega)=C_{-}^{+}(\omega)$). When $n$ is even
\cite{AS3}, one finds that $C_{-}^{-}(\omega)=1$ and
$C_{-}^{+}(\omega)=0$. This corresponds to the physical statement
that there is no particle creation (no mode mixing) in dS space
for an odd number of spacetime dimensions. For $n$ odd, there
is nontrivial mode mixing with,
\beq
\labell{deven}
C_{-}^{-}(\omega) = \coth \pi \omega \;\;\;\;\;\; C_{-}^{+}(\omega) = (-1)^{\frac{n+1}{2}}
\frac{1}{\sinh \pi \omega},
\eeq
where $|C_{-}^{-}(\omega)|^{2}-|C_{-}^{+}(\omega)|^{2}=1$ holds
since the modes are properly normalized throughout their
evolution. The expression for the phase in eq.~(\ref{bogprinc}) is
\beq
\labell{phase}
e^{-2i\theta_{L}} = (-1)^{L-\frac{n}{2}} \frac{\Gamma(-i\omega)\Gamma(L+\frac{n}{2}+i\omega)}
{\Gamma(i\omega)\Gamma(L+\frac{n}{2}-i\omega)}.
\eeq
It is clear that for large enough $\omega$, the mixing coefficient
$C_{-}^{+}(\omega)$ becomes negligible which is in accord with the
intuition that there will be limited particle production in high
energy modes. We will find that in the other two series there is
no phase comparable to eq.~(\ref{phase}). This complicates the
expressions for mode mixing between the boundaries and will lead
to interesting features.

\subsection{Complementary series and tachyonic fields}

For the modes of both the complementary and the tachyonic series,
the weights $h_{+}$ and $h_{-}$ are real and so the above mode
functions are entirely real,
\beq
\labell{ct1}
y_{L+}(t) = \bar{N}_{+} \cosh^{L}\!t\, e^{(L+\frac{n}{2}+\mu)t}
F(L+\frac{n}{2},L+\frac{n}{2}+\mu;1 + \mu;-e^{2t}),
\eeq
\beq
\labell{ct2}
y_{L-}(t) = \bar{N}_{-} \cosh^{L}\!t\, e^{(L+\frac{n}{2}-\mu)t}
F(L+\frac{n}{2},L+\frac{n}{2}-\mu;1- \mu;-e^{2t}).
\eeq
Hence with respect to the usual Klein-Gordon product
these two solutions have zero norm, \ie $(\phi_{L+},\phi_{L+})=0=(\phi_{L-},\phi_{L-})$.

Of course, this is not unnatural. One gains intuition by considering
the usual plane wave decomposition in flat
spacetime. There, one may choose between two bases, the one involving complex exponentials
and the one involving {cosines} and {sines}. The latter basis in fact has the same
characteristics as the present modes in the complementary series
in terms of normalization with respect to the Klein-Gordon inner product.
Consequently, to define a reasonable normalization for the mode functions (\ref{ct1})
and (\ref{ct2}), we require $(\phi_{L-},\phi_{L+})=i$
\beq
\bar{N}_{+} =\frac{2^{L+\frac{n-1}{2}}}{\sqrt{\mu}} =
\bar{N}_{-},
\eeq
where we have resolved the remaining ambiguity by simply demanding that
$\bar{N}_{+} =\bar{N}_{-}$.
With this choice of normalization, it is clear that upon quantizing the
scalar field in the dS background the corresponding mode coefficients will have
commutation relations analogous to those of coordinate and momentum operators, rather than
raising and lowering operators.

As in the previous subsection, by substituting $t\rightarrow-t$, we define modes
$y_L{}^\pm(t)\equiv y_{L\pm}(-t)$ which have the simple time dependence of
eq.~\reef{simpletime} in the asymptotic region approaching $I^+$.
Using a simple identity of hypergeometric functions \cite{handbook},
one can relate the two sets of modes as
\beqa
y_{L-}(t) &=& \bar{C}_{-}^{-}(\mu)\, y_{L}{}^{-}(t) + \bar{C}_{-}^{+}(\mu)\, y_{L}{}^{+}(t),
\nonumber\\
\labell{defcon6}
y_{L+}(t) &=& \bar{C}_{+}^{-}(\mu)\,
y_{L}{}^{-}(t) + \bar{C}_{+}^{+}(\mu)\, y_{L}{}^{+}(t),
\eeqa
where the elements of the mixing matrix $C$ (the Bogolubov
coefficients) are given by
\beq
\labell{yikes1}
\bar{C}_{-}^{+}(\mu) = \frac{\Gamma(1-\mu)\Gamma(-\mu)}{\Gamma(\frac{2-n}{2}-\mu-L)
\Gamma(\frac{n}{2}-\mu+L)},\;\;\;\;
\bar{C}_{-}^{-}(\mu) = -(-1)^{L} \frac{\sin (\frac{\pi\,n}{2})}{\sin \pi \mu},
\eeq
\beq
\labell{yikes2}
\bar{C}_{+}^{-}(\mu) = -\frac{\Gamma(1+\mu)\Gamma(\mu)}{\Gamma(\frac{2-n}{2}+\mu-L)
\Gamma(\frac{n}{2}+\mu+L)},\;\;\;\;
\bar{C}_{+}^{+}(\mu) = -(-1)^{L}\frac{\sin (\frac{\pi\,n}{2})}{\sin \pi \mu}.
\eeq
We now describe some features of the resulting mode mixing for the
complementary series. In this case, recall that $0<M^2\le n^2/4$
which implies that $0\le\mu<n/2$. Of course, certain features
depend on the spacetime dimension $n$+1 as before:

\noindent{\it a) $n$ odd:} Generically for the case of an even
spacetime dimension, there is nontrivial mode mixing. An exception
occurs for $\mu=(2m+1)/2$ with $m$, a positive integer. For these
special cases, there is no mixing since
$\bar{C}_{-}^{+}=0=\bar{C}_{+}^{-}$ and one finds that
$\bar{C}_{-}^{-}=\bar{C}_{+}^{+}=(-1)^{{n\over2}+\mu}(-1)^L$.

\noindent{\it b) $n$ even:} Generically for this case of an odd
number of spacetime dimensions, one finds
$\bar{C}_{+}^{+}=0=\bar{C}_{-}^{-}$ and
$\bar{C}_{-}^{+}\,\bar{C}_{+}^{-}=-1$ (where $\bar{C}_{-}^{+}$ and
$\bar{C}_{+}^{-}$ both have a nontrivial dependence on $L$). This
means that a mode that is scaling like $e^{h_{\pm}t}$ on $I^{-}$
will have the `opposite' scaling $e^{-h_{\mp}t}$ on $I^{+}$. We
refer to this phenomenon as `maximal mixing'. This phenomenon is
absent when $\mu$ is an integer. This case must be treated with
some care as the solution for $y_{L-}$ appearing in eq.~\reef{ct2}
breaks down.\footnote{Similar remarks apply for $n$ odd and $\mu$
integer, but in that case one still finds nontrivial mode mixing.}
The correct solution \cite{handbook} has an additional logarithmic
singularity near $I^-$, \ie, subdominant power law behavior in
$t$. In any event, the final result for $n$ even and $\mu$ integer
is: $\bar{C}_{-}^{-}=\bar{C}_{+}^{+}=(-1)^{{n\over2}+\mu+1}(-1)^L$
and $\bar{C}_{-}^{+}=0=\bar{C}_{+}^{-}$.

Finally we briefly consider the tachyonic or discrete series
\cite{RLN,RLN2,RLN3}. Recall that in this case with $M^2<0$,
$h_-<0$ so that the modes scaling as $e^{\pm h_{-}t}$ diverge as
one approaches either $I^-$ or $I^+$, respectively. Generically
there is nontrivial mode mixing and so even if a mode is
convergent at one asymptotic boundary it will be divergent at the
opposite boundary. However, an interesting exceptional case is
when a $y_{L+}$ mode (scaling like $e^{h_+t}$ as $t\rightarrow
-\infty$) evolves to the corresponding $y_L{}^+$ mode (with
$e^{-h_+t}$ behavior for $t\rightarrow\infty$). Such a mode would
have convergent behavior both towards the future and past
boundaries. This behavior would result when $\bar{C}_{+}^{-}$
vanishes. A brief examination of eq.~\reef{yikes2} requires that
$1+|h_-|-L$ is zero or a negative integer. As above this
constrains $\mu$ to be an integer or half-integer depending on the
spacetime dimension. We express this constraint in terms of the
(tachyonic) mass
\beq
\labell{piker}
-M^2=\left\lbrace
\matrix{\frac{1}{4}((2m+1)^2-n^2)&{\rm for}\ n{\ \rm odd}&{\rm with}\ m=(n-1)/2,(n+1)/2,
\ldots\cr
m^2-\frac{n^2}{4}&{\rm for}\ n\ {\rm even}&{\rm with}\ m=n/2,n/2+1,\ldots\hfill\cr}
\right.
\eeq
This constraint gives rise to the nomenclature that these modes
are often referred to as the discrete series. However, the above
constraint is not sufficient; rather we must also impose a
constraint on the `angular momentum' quantum number $L$, namely,
\beq
L\ge 1+|h_-|\ .
\eeq
Hence the completely convergent modes only appear for sufficiently
large angular momenta. Note that it is still true that using the
usual Klein-Gordon inner product these modes have a vanishing norm
$(\phi_{L+},\phi_{L+})=0$. However, in the mathematics literature
(\eg \cite{RLN3}) these modes are singled out by having finite
norm in the sense given by the spacetime integral: $\int d^{n+1}x
\sqrt{-g}\,|\phi_{L+}|^2=1$.

This construction shows that even in the tachyonic mass range, one can
find certain normalizable modes (in the above sense) for special choices
of parameters. However, we
reiterate that
while these formal results for the discrete series may be interesting mathematically, they
are not useful in understanding the physics of dS space. As emphasized above in the
discussion
of the dS/CFT correspondence, one must consider the full space of solutions, and presently
even in the exceptional cases, the normalizable modes are accompanied by modes diverging
at both asymptotic boundaries.  Thus, the normalizable modes do not form
a complete set of modes on a Cauchy surface. Such divergences, which occur in the generic case as well,
are simply an indication that a linearized analysis of tachyonic fields is inappropriate.
Of course, nonlinear field theories with potentials including unstable (or metastable)
critical points may play an important role in the paradigm of inflationary cosmology, and
such theories can produce interesting asymptotically dS spacetimes \cite{tale1}. Our
point here is simply that one should consider the full nonlinear evolution of such fields
including their backreaction on the spacetime geometry.

\section{Stability of scalar modes}
\label{appendix3}

Here we discuss the stability of dS space with respect to the
various scalar field modes introduced in appendix \ref{appendix1}.
In ref.~\cite{rgflows2}, an attempt is made to distinguish the
modes associated with $\tphi_{0\pm}$ boundary data on the basis of
an `energy functional.' However, one might find this result
unsatisfactory given that the `energy' is not conserved and so it
does not give a covariant indicator by which we might measure
backreaction effects. So in the following we readdress the
question of whether or not these scalar mode functions can
successfully be regarded as fluctuations without taking into
account any backreaction effects. Our conclusion below is that for
the principal or complementary series (\ie $M^{2}>0$), all of the
modes can consistently introduce only a small perturbation
throughout the entire time evolution, including the asymptotic dS
regions. Therefore, the $\tphi_{0\pm}$ modes cannot be
distinguished on this basis. Similar arguments were previously
sketched out in ref.~\cite{Bousso:2002fi}.

To begin, consider a free scalar field propagating in a fixed dS
background. In the asymptotic future, \ie $t\rightarrow+\infty$,
the background metric \reef{metricds} takes the form
\beq ds^2\simeq-dt^2+{1\over4}e^{2t/\lc}\,d\Omega_n^2 .
\labell{metric} \eeq
As discussed above, in this asymptotic region, the solutions
of scalar field equation \reef{eomfirst} take the form
\beq \phi\simeq e^{-h_\pm t/\lc}\,\tphi_{0\pm}(\Omega),
\labell{form} \eeq
where as given in eq.~\reef{water},
$h_\pm={n/2}\pm\sqrt{{n^2/4}-M^2\lc^2}$. Note that we are assuming
that the spatial wavefunctions $\phi_\pm(\Omega)$ have reasonable
behavior on the $n$-sphere, but the precise details will be
unimportant.

Consider the contribution of the scalar to the right-hand-side of
Einstein's equations:
\beq \labell{eineq} R_{\mu\nu}-{1\over2}g_{\mu\nu}R=8\pi
G_{{\scriptscriptstyle N}}\, T_{\mu\nu}(\phi)-\Lambda\,g_{\mu\nu}\
. \eeq
Taking $\phi$ to be a complex scalar field for simplicity, the
scalar stress tensor is
\beq
T_{\mu\nu}=2\nabla_\mu\phi^*\nabla_\nu\phi-g_{\mu\nu}\left(
g^{\sigma\rho}\,\nabla_\sigma\phi^*\nabla_\rho\phi+M^2|\phi|^2\right)\ .
\labell{stress} \eeq
Now to address the stability of the asymptotic dS geometry in
eq.~\reef{metric}, we will compare the above source in Einstein's
equations coming from the scalar field with that arising from the
cosmological constant, which defines the background geometry.

In general, one might look at the components of the stress tensor
in an orthonormal frame so that they will correspond to the
physical energy density and stresses observed by an inertial
observer. Note, however, in the metric \reef{metric} above, the
coordinate time $t$ really is the proper time of comoving
observers and so by focussing on the energy density $T_{tt}$, one
need not worry about transforming between coordinate and frame
indices. Note that the analysis of the pressure components yields
the same results as described for the energy density below.
Proceeding from eq.~\reef{stress}, the relevant energy density is
\beq
T_{tt}=|\prt_t\phi|^2+|\hat\nabla_k\phi|^2+M^2|\phi|^2\ .
\labell{energy}
\eeq
For comparison purposes, we also consider the contribution of the
cosmological constant to Einstein's equations \reef{eineq}. The
corresponding energy density is a constant:
\beq
T_{tt}=\Lambda={n(n-1)\over2\lc^2}\ .
\labell{cosmo}
\eeq

Let us begin by considering scalar mass parameters corresponding
to the complementary series, \ie $0<M^2\lc^2<n^2/4$, so that the
exponents $h_\pm$ are real. With the wavefunctions given in
eq.~\reef{form}, the energy density \reef{energy} becomes
\beqa
T_{tt}&\simeq& {1\over \lc^2}\left(h_\pm^2+M^2\lc^2\right)
e^{-2h_\pm t/\lc}\, |\tphi_{0\pm}|^2+O\left(e^{-2(h_\pm-1)
t/\lc}\right)
\nonumber\\
&\simeq&{1\over \lc^2}\left({n^2\over2}\pm
n\sqrt{{n^2\over4}-M^2\lc^2} \right) e^{-2h_\pm
t/\lc}\,|\tphi_{0\pm}|^2 \labell{energy1} \eeqa
Note that the spatial gradients in the stress tensor \reef{energy}
make subleading contributions above. Now this expression shows
that the contribution of the scalar to the local energy density
decays in the asymptotically dS region. Certainly this criterion
only holds for the complementary series with $0<M^2\lc^2<n^2/4$.
When the mass is in this range, the perturbation which these modes
introduce in the Einstein equations becomes diminishingly small in
the asymptotic region. Of course, this result matches the naive
intuition that one might derive from the simple observation that
the mode functions are decaying as $t\rightarrow\infty$. In any
event, if the scalar fluctuations begin as small perturbations,
they remain a `small' disturbance throughout the evolution of the
spacetime (and the scalar field). Therefore this analysis confirms
that the perturbative treatment of the scalar field is consistent
and that the asymptotic dS geometry is stable against the
introduction of such perturbations.

Similarly, we may consider mass parameters in the range of the
principal series, \ie for $M^2\lc^2>n^2/4$. Of course, in this
case, the exponents are complex: $h_\pm=-n/2\pm i\omega$. Now, the
scalar energy density \reef{energy} becomes
\beqa
T_{tt}&\simeq&\left({|h_\pm|^2\over L^2}+m^2\right)
e^{-n\,t/\lc}|\tphi_{0\pm}|^2+\cdots
\nonumber\\
&\simeq&2m^2e^{-n\,t/\lc}\,|\tphi_{0\pm}|^2\ .
\labell{energy2}
\eeqa
Hence we see that in this mass range, in accord with naive
expectations, the disturbance of scalar modes to Einstein's
equations is small. As above then, we conclude that the asymptotic
dS geometry is stable these perturbations.

For a tachyonic field with $M^2<0$, the exponents are real but
$h_-$ is negative. Therefore the corresponding scalar
perturbations are growing exponentially in the asymptotic dS
region. The asymptotic scalar energy density is again given by the
expression in eq.~\reef{energy1}. Hence we see that the
contribution of the $\tphi_{0-}$ modes to the local energy density
is growing without bound in this region of the spacetime.
Therefore the energy density of the scalar field would quickly
overwhelm that of the cosmological constant \reef{cosmo}
irrespective of how small the perturbations began. Of course, all
we can really conclude is that with the growth of these modes, the
system will enter a nonlinear regime where the scalar field can no
longer be consistently treated using a linearized perturbation
analysis. In any event, we interpret this result as indicating the
exponential growth of these (or any tachyonic) fields produces an
instability as one cannot expect the asymptotic spacetime geometry
to resemble dS space \reef{metric}. This confirms the naive
expectations for tachyonic fields in dS space. This result is in
complete agreement with the previous discussion stating that a
successful analysis of such fields must consider the full
nonlinear evolution of the scalar including its backreaction on
the spacetime geometry.

To reiterate our conclusions, let us note that the analyses here
and in appendix \ref{appendix1} apply to global coordinates on dS
space. Further the explicit mode functions for the principal and
complementary series given in appendix \ref{appendix1} are
well-behaved and bounded throughout the entire spacetime.
Therefore in these cases with suitably `small' mode coefficients,
the above analysis indicates that the scalar energy density
\reef{energy} will be negligible compared to that introduced by
the background cosmological constant throughout the spacetime.
Hence in these cases, the scalar field will provide a small
perturbation throughout the entire evolution of the dS spacetime,
not only in the asymptotic regions. It is also clear that this
result applies for both the $\tphi_{0\pm}$ modes and so these two
distinct boundary data cannot be distinguished on this basis.


\end{document}